\documentclass[a4paper]{jpconf}

\usepackage{graphicx}
\begin{document}
\title{Heater induced thermal effects on the LTP dynamics}

\author{F Gibert$^{1,3}$, A Lobo$^{1,2}$,
M Diaz-Aguil\'o$^{1,4}$, I Mateos$^1$, M Nofrarias$^5$, J Sanju\'an$^6$, 
A Conchillo$^{1,2}$, L Gesa$^{1,2}$ and I Lloro$^{1,2}$}
\address{$^1$ Institut d'Estudis Espacials de Catalunya (IEEC), Barcelona, Spain}
\address{$^2$ Institut de Ci\`encies de l'Espai (CSIC-IEEC), Barcelona, Spain}
\address{$^3$ Escola T\`ecnica Superior d'Enginyeries Industrial i Aeron\`autica de Terrassa (ETSEIAT), Universitat Polit\`ecnica de Catalunya (UPC), Terrassa, Spain}
\address{$^4$ Escola Polit\`ecnica Superior de Castelldefels (EPSC), Universitat Polit\`ecnica de Catalunya (UPC), Castelldefels, Spain}
\address{$^5$ Max-Planck-Institut f\"ur Gravitationsphysik, Albert Einstein Institut (AEI), Hannover, Germany}
\address{$^6$ Department of Physics, University of Florida, Gainesville, FL, United States}

\ead{gibert@ieec.cat}

\begin{abstract}
The STOC (Science and Technology Operations Centre)
simulator of the LPF (LISA PathFinder) mission is intended to
provide a validation tool for the mission operations tele-commanding
chain, as well as for a deeper understanding of the underlying physical
processes happening in the LTP (LISA Technology Package). Amongst the different physical
effects that will appear onboard, temperature fluctuations in the Electrode Housing (EH)
could generate disturbances on the interferometer (IFO) readouts, therefore
they must be known and controlled. In this
article we report on the latest progress in the analysis at IEEC
of the LTP response to thermal signals injected by means of heaters.
More specifically, we determine the transfer functions relating
heat input signals to forces on the Test Masses (TMs) in the
LTP frequency band, from 1\,mHz to 30\,mHz. A complete thermal model
of the entire LPF spacecraft plus payload, elaborated and maintained
at European Space Technology Center (ESTEC), was used to obtain temperature distributions in response to
heat inputs at prescribed spots (heaters), which are later processed
to calculate the associated dynamical effects on the Test Masses.
\end{abstract}

\section{Introduction}

Undesired temperature fluctuations around the TMs and in other points of LTP
have been identified as disturbances that
could alter IFO readouts. In order to keep them controlled, a set of experiments, gathered in the
Experiment Master Plan (EMP), are planned to
be conducted onboard LPF so the effects can be measured and later use the results to assess the
weight of temperature noise in the global noise budget.

All this involves the necessity of modelling and simulating the different thermal effects. 
To achieve this goal, the STOC simulator of LPF, that is aimed to recreate the behavior
of the spacecraft through its main operational phases, must include them.

In this article, we focus only on the consequences of thermal effects at EH due to the 
activation of the different 
Data and Diagnostics System (DDS) heaters following the plans from the EMP. Specifically, 
in the following a description of the
process considered to determine forces acting on TMs is explained, as well as the simulations of the heat input
in the system. In the last sections, results from the simulations are presented and discussed.

\section{Basis}

\subsection{General layout of DDS thermal items}
\begin{figure}
  \begin{center}
    \includegraphics[width=27pc]{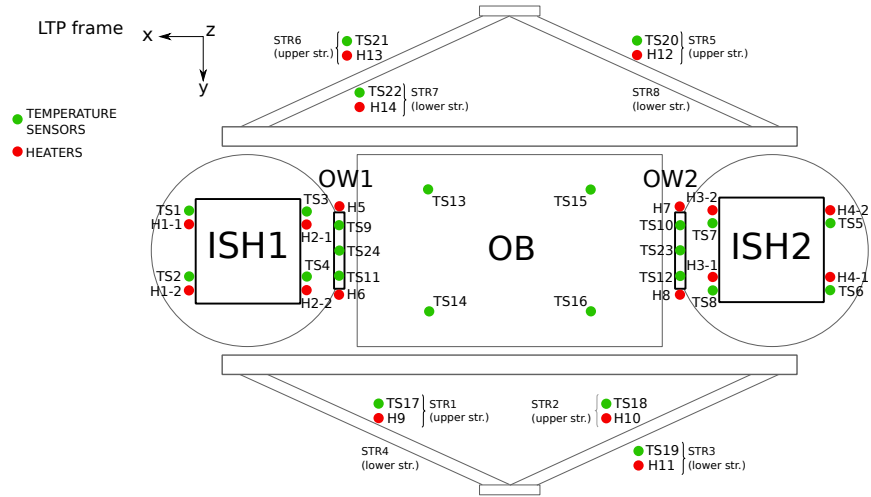}
    \caption{\label{global_layout}Figure caption for first of two sided figures.}
  \end{center}
\end{figure}

\begin{figure}[h]
\begin{minipage}{21pc}
  \includegraphics[width=21pc]{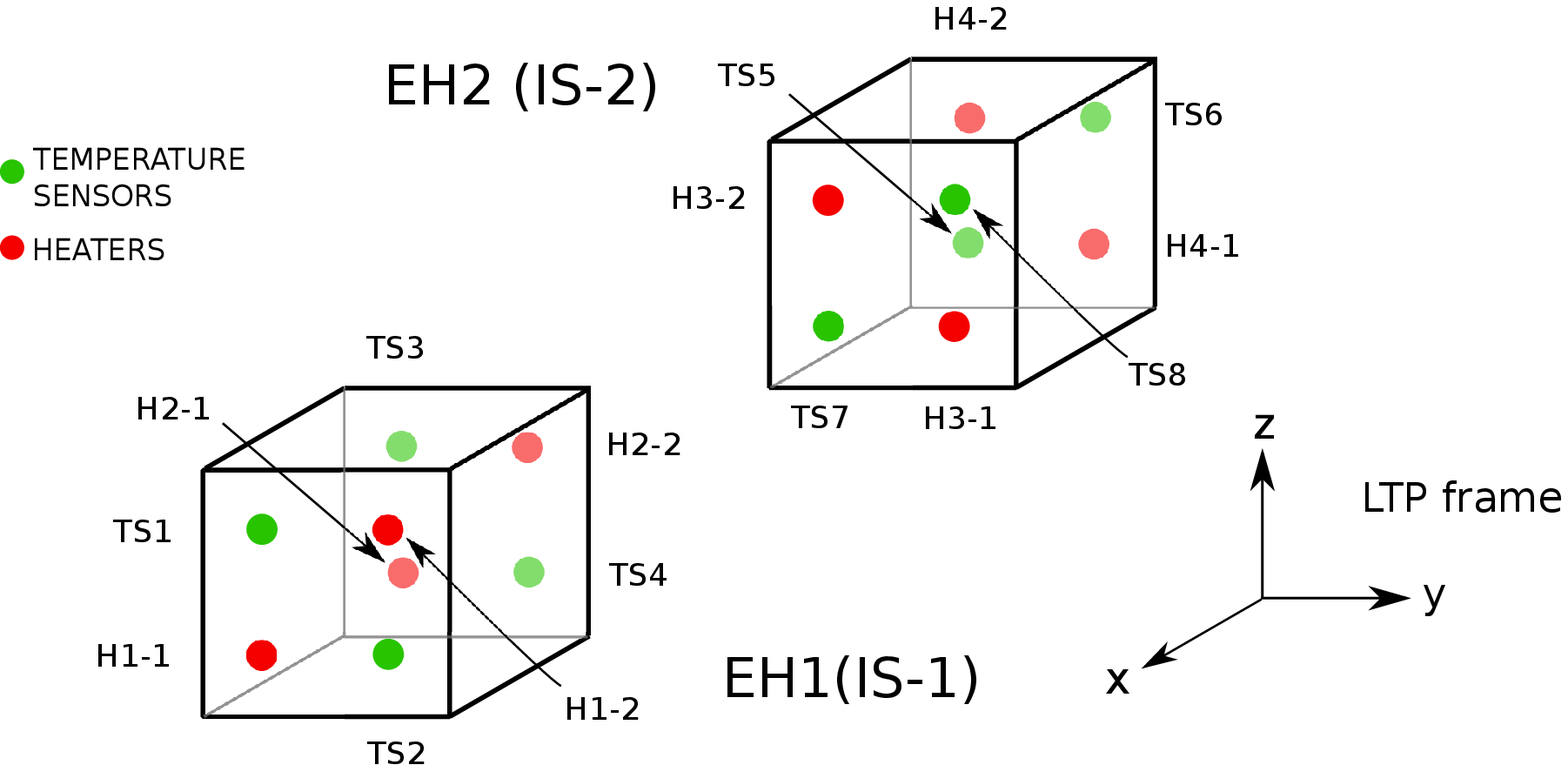}
\caption{\label{eh_layout}Detail of EH DDS items.}
\end{minipage}\hspace{2pc}
\begin{minipage}{14pc}
  \includegraphics[width=14pc]{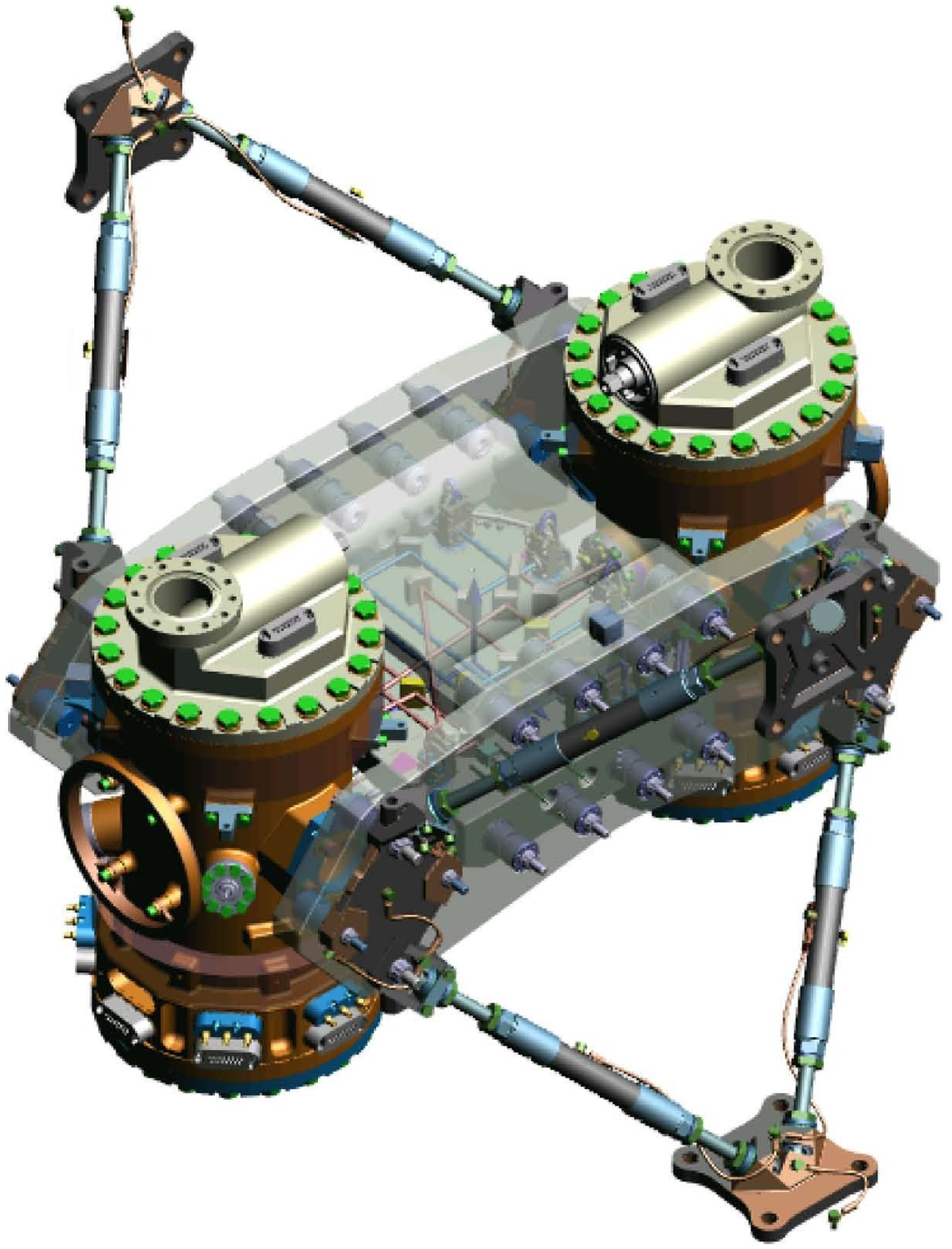}
\caption{\label{ltp_picture}General picture of the LTP.}
\end{minipage} 
\end{figure}

The DDS thermal items are composed by 18 {\it physical} heaters and 24 temperature
sensors \cite{layout}. The general scheme of the distribution and numenclature is shown in Figure~\ref{global_layout},
and a scheme only of the EH is presented in Figure~\ref{eh_layout}. Figure~\ref{ltp_picture}
shows a global picture of the LTP.

\subsection{EMP thermal calibration description}
The Experiment Master Plan (EMP) thermal calibration aims to characterize the 
consequences of assymetric temperature fluctuations happening in the LTP.
Experiments consist of applying well-defined heat loads at specific points of LTP and 
observe induced temperature variations at particular points in order to see which are their consequences.
Their ultimate goal is to calibrate these effects so that the temperature noise can eventually
be subtracted from the IFO readout.

Thermal EMP exercises can be divided into three groups, depending on the main focus of the
exercise and the heaters that are being activated:

\newcounter{alp}
\newcommand{\abc}{\item[(\alph{alp})]\stepcounter{alp}}
\newenvironment{liste}{\begin{itemize}}{\end{itemize}}
\newcommand{\aliste}{\begin{liste} \setcounter{alp}{1}}
\newcommand{\zliste}{\end{liste}}
\newenvironment{abcliste}{\aliste}{\zliste}

\begin{abcliste}
 \abc Determination of forces and torques on the TMs due to EH heaters activation.
 \abc Determination of IFO phase shifts due to Optical Window (OW) heaters activation.
 \abc Determination of IFO phase shifts due to Suspension Struts (STR) heaters activation.
\end{abcliste}

Dynamic effects from exercises performed at {\it (a)} come as a result of the existence of
different thermal effects appearing in the EH environment - see Section~\ref{thermeff}-, while 
experiments from {\it (b)} and {\it (c)} are aimed to analyze thermoelastic effects
caused by heating the OWs and creating stresses on the Optical Bench, respectively, and should not 
interfere with the TMs dynamics.

The specific input power sequences are described in Table~\ref{EMP_table} \cite{EMP, inputheat}. 
They are intended to generate
enough {\it disturbance} in specific spots so the IFO can detect their presence. However, they may present
side effects that must be also studied, particularly in the dynamics of the TM during 
exercises {\it(b)} and {\it(c)}, due to the amount of total heat applied.

\begin{table}[h]
\caption{\label{EMP_table}Input signals for each experiment. Experiments from {\it(a)} apply heat 
 to pairs of EH heaters from same face of a TM alternatively, while in ones from {\it(b)} and
{\it(c)} simple pulse series are applied to single a heater each time.}
\begin{center}
\begin{tabular}{lllll}
\br
Experiment & Power & Period & Duty cycle & Duration \\
\mr
(a) EH & 1-80\,mW & 2000\,s & 50\% & 4000\,s\\
(b) OW & 0.1-1\,W & 1000\,s & 5\% & 5000\,s\\
(c) STR & 0.5-2\,W & 2500\,s & 4\% & 7500\,s\\
\br
\end{tabular}
\end{center}
\end{table}

\subsection{Signal requirements}

LPF main noise requirements are given as a spectral density of noise by \cite{signalreqs}
  \begin{eqnarray}
    S^{1/2}_{\Delta a,\, {\rm LPF}}(\omega)\leq 3\times10^{-14}\left[1+\left(\frac{\omega/2\pi}{3\, {\rm mHz}}\right)^{2}\right]\, {\rm m}\, {\rm s}^{-2}\, {\rm Hz}^{-1/2}
    \label{noise}
  \end{eqnarray}
in the band from 1mHz to 30\,mHz. Considering that the mass of each TM
is 1.96\,kg \cite{EPB}, Equation~\ref{noise} consequent conversion to force requirement
is expressed with
\begin{eqnarray}
    S^{1/2}_{\Delta F,\, {\rm LPF}}(\omega)\leq 5.9\times10^{-14}\left[1+\left(\frac{\omega/2\pi}{3\, {\rm mHz}}\right)^{2}\right]\, {\rm N}\,  {\rm Hz}^{-1/2} 
    \label{noise_force}
\end{eqnarray}
On the other hand, the SNR requirement to detect thermally created signals in the inertial sensor (IS)
with the IFO is $SNR\geq50$ \cite{signalreqs},
and it is obtained, considering a force signal $F(t)$, as
\begin{eqnarray}
    (SNR)^2= \frac{2}{\pi} \int_{BW} \frac{|\tilde{F}(w)|^2}{S_{\Delta F,\, {\rm LPF}}(w)}dw
    \label{snr}
\end{eqnarray}
where $\tilde{F}(w)$ is the Fourier Transform of the force signal $F(t)$.

\subsection{Thermal effects}
\label{thermeff}
Temperature differences across the TMs environment can produce differential pressures that turn 
to net forces and torques on the TMs. Three different thermal effects have been identified
as mechanisms responsible of these quantities \cite{miquelsthesis}:

\subsubsection{Radiation pressure effect} 
It is based on the temperature-dependence of the 
radiation emitted by a heated surface. The pressure generated is expressed as:
\begin{eqnarray}
    P_{\rm rdn}=\frac{8}{3}\frac{ \sigma T^{3}}{c}\Delta T\;\;{\rm Pa} 
    \label{eq.1n1}
\end{eqnarray}
where {\it $\sigma$} is the Stefan-Boltzmann constant, {\it c} is the speed of light,
$T$ is the absolute temperature and $\Delta T$ is the temperature difference between two surfaces.
\subsubsection{Radiometer effect} 
It appears in rarified atmospheres where the particles have 
a mean free path much longer than the distance between the two surfaces. The consequent pressure is 
represented here as:
\begin{eqnarray}
    P_{\rm rmt}=\frac{1}{4}\frac{p}{T}\Delta T\;\;{\rm Pa}
    \label{eq.1n2}
\end{eqnarray}
where {\it p} stands for the remaining gas static pressure. $T$ and $\Delta T$ have same
meanings as in Equation~\ref{eq.1n1}.

\subsubsection{Outgassing effect} A third thermal effect with poorly known consequences, the outgassing effect
could appear. Models of this effect made so far
are not accurate enough, therefore it will not be considered here.

\section{Procedure}

The procedure can be split in two parts, the first one concerning all the operations related 
with the data from the thermal model and the following fit that allows to calculate the
temperature distribution, and the second one related with the force modelling.

  \subsection{Thermal model}
The thermal model used to simulate the thermal behaviour of the different parts under analysis is the current
ESATAN version that can be found at ESTEC.
Samples of transfer functions containing the temperature response
in a node after applying heat into another can be directly determined by the ESATAN program.

A set of specific thermal nodes concerning the temperature sensor nodes and the nodes
of the different electrodes in the EH were selected as representative of the temperature
distribution response \cite{nodes}, and their transfer function samples were provided by ESTEC.
Data was fitted using LTP Data Analysis Tool software in order to get analytic
expressions for each {\it node to node} relation.

Additionally to these transfer functions, specific thermal behavior for the heaters modelling
their thermal resistance was considered also \cite{pepsthesis}, and it was combined with the initial transfer
functions so as to get a more approximate model of the case. With the definitive relation, time response of temperature
at different points due to the activation of different heaters were directly determined.

\begin{figure}[h]
\begin{minipage}{0.69\textwidth}
\includegraphics[width=24pc]{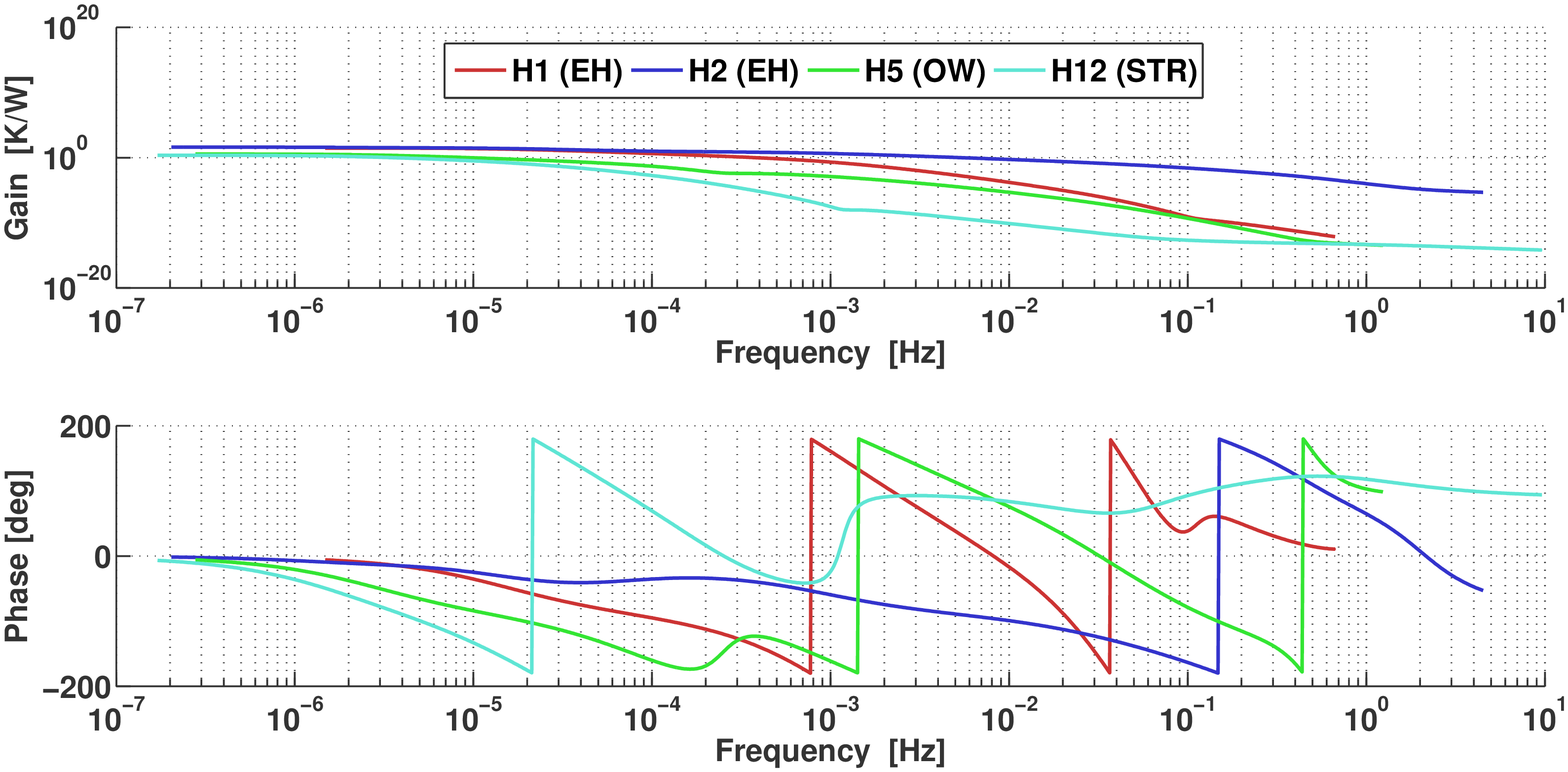}
\end{minipage}
\begin{minipage}{0.29\textwidth}\caption{\label{bode_ex}Bode plots of the temperature response at
temperature sensor TS3 location (-X face of EH1). The 
different heaters considered are H1, H2, H5 and H12.}
\end{minipage}
\end{figure}

Figure~\ref{bode_ex} shows a group of Bode Plots as an example of the different transfer functions obtained.

  \subsection{Modelling forces}
Considering the geommetry of the case and assuming that only the faces containing EH heaters are relevant
for the calculation of forces along the X axis, it is possible to express the resulting force with Equation~\ref{eq_res_force}.
\begin{equation}
 F_X=F_{\rm rdn}+F_{\rm rmt}=\frac{8}{3}\frac{\epsilon_{ij}A\sigma T_0^{3}}{c}\Delta T_{AB}+\frac{1}{4}\frac{Ap}{T_0}\Delta T_{AB}=\left(\frac{8}{3}\frac{\epsilon_{ij}A\sigma T_0^{3}}{c}+\frac{1}{4}\frac{Ap}{T_0}\right)\Delta T_{AB}\;\;{\rm N}
\label{eq_res_force}
\end{equation}
where $A$ is the area where the pressure is applied and $\epsilon_{ij}$ is the view
factor between two EH-TM surfaces.

Replacing with the nominal values of the case \cite{EPB} and considering a view 
factor $\epsilon_{ij}=0.683$,
the resulting temporal {\it force-temperature} relation
states $F_X(t)=36.4\Delta T(t) \,\,{\rm pN}$,
where $\Delta T(t)$ is the temperature difference between both X inner faces of a single EH at
time $t$.

\section{Results}
  \subsection{EH heaters}

\begin{figure}[h]
\begin{minipage}{17pc}
  \includegraphics[width=17pc]{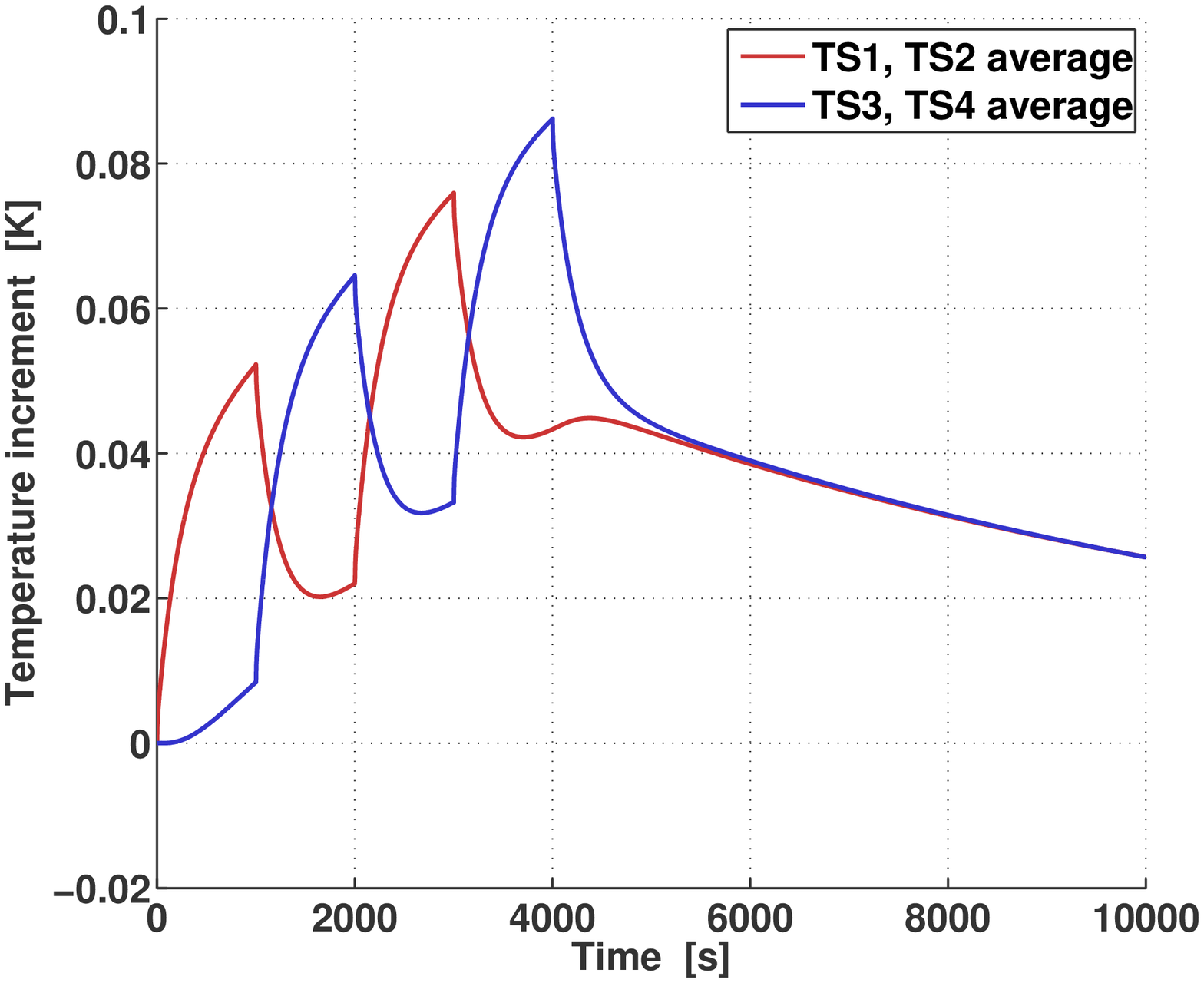}
\end{minipage}
\hspace{3pc}
\begin{minipage}{17pc}
  \includegraphics[width=17pc]{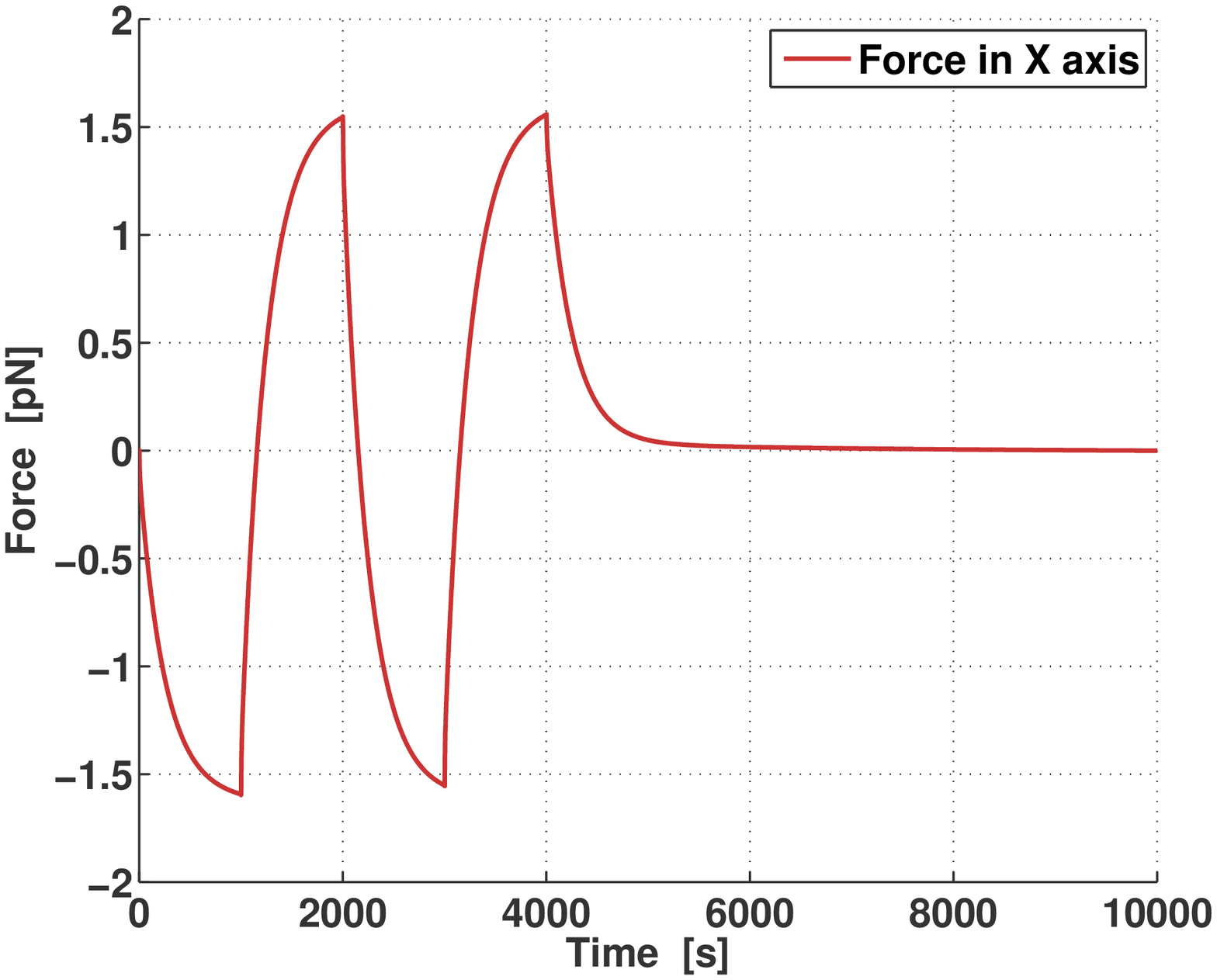}
\end{minipage}
\begin{minipage}{17pc}
\caption{\label{label1}Absolute temperature increment values for each side of TM-1 along exercise {\it (a)}.}
\end{minipage}
\hspace{3pc}
\begin{minipage}{17pc}
\caption{\label{label2}Force in X axis on TM-1 along exercise {\it (a)}. \\ \; }
\end{minipage}
\end{figure}

Figures~\ref{label1}~and~\ref{label2} present the time response of the temperature gradient and force in X axis with
a power input of 10mW. The higher amplitude of the first peak in Figure~\ref{label2} is created due to the
initial {\it cold} environment in the EH. Therefore, the total force applied will tend to be assymetrical. 

The SNR obtained from this exercise is of about 116, considering an integration time of 50,000 seconds, 
what meets the requirement. 

  \subsection{Side effects on EH}

Exercise {\it (b)} produces a SNR due to forces in EH of $\sim$0.93 when a power of 0.2\,W is applied, while exercise {\it (c)} gets
a SNR of $\sim$0.02 with a power of 2\,W. In both cases, an integration time of 100,000 seconds was considered. 

  \subsection{Transient times}
The transient time length is also of great interest, because it sets the required relaxation time between
different exercises and the total length of the experiment. Taking into account that considerable amounts of power are applied 
in some cases, it should be expected that long periods of time would be required between different exercises. 

For example, referring to exercise {\it (a)}, the time that it takes for the EH to cool back to
10\% of the maximum absolute temperature reached is about 22,000\,s, while the time that it needs to recover to 10\% of
maximum differential temperature is about 4,600\,s.

However, for the cases of exercises {\it (b)} and {\it (c)} it is not that clear, as the hottest points are reached
far from the EH. Relaxation times of
these exercises can easily be a matter of days, although they vary with the distance of the heater to each 
EH and the total amount of heat applied.

\section{Conclusions}
Thermal response of the EH thermistors to the activation of the different LTP heaters already analysed.
Analysis results so far point to:
 \begin{itemize}
  \item EH heaters activation parameters produce forces on the TM which are 100 or more times the background noise.
  \item Depending on the applied power, OW heating can be sensed in the EH with SNR around 1.
  \item STR heating does not cause noticeable effects at TM. 
 \end{itemize}
It is important to remark that the current power margins allow a set of possible heat inputs that in some
cases could involve quite long relaxation times. In order to assess them appropiately, more detailed analysis 
of each case should be done, including thermoelastic effects, which are beyond the scope of this paper.
Further work is still required, specifically in detailing a more accurate temperature distribution
around the TMs, that will be obtained from ESATAN-provided electrode temperatures. This more detailed data will
have to be correlated with the EH temperature sensor readouts.

Finally, forces in the Y and Z axes of the EH must, and will be studied, as well as torques
generated from the same sources, although their impact is not expected to affect too much. Work is currently
underway on these matters.
 
\ack This study has been supported by MICINN-CSIC contract ESP2007-61712. FG also acknowledges a grant 
from the project {\it Hibridaci\'o d'assignatures d'elevat 
contingut tecnol\`ogic com a prova pilot per als nous plans d'estudis de Grau i M\`aster
de la titulaci\'o d'Enginyeria Aeron\`autica a l'ETSEIAT}, from “Departament
de Projectes d'Enginyeria” at ETSEIAT-UPC.


\section*{References}
\begin{thebibliography}{9}
\bibitem{layout} G Kahl 2007, {\it DDS Subsystem Specification}, Report S2-ASD-RS-2004, v4.4.
\bibitem{EMP} M Nofrarias {\it et al} 2009, {\it Thermal experiments on board LTP}, Report S2-IEC-TN-3042, v1.1.
\bibitem{inputheat} M Nofrarias 2006, {\it DDS heaters characterisation}, Report S2-IEC-TN-3025, v1.1.
\bibitem{EPB} N Brandt {\it et al} 2007, {\it Experiment Performance Budget}, Report S2-ASD-RP-3036, v2. 
\bibitem{signalreqs} S Vitale {\it et al} 2005, {\it Science Requirements and Top-Level Architecture
Definition for the LISA Technology Package (LTP) on Board LISA Pathfinder(SMART2)}, Report LTPA-UTN-ScRD, Iss3-Rev1. 
\bibitem{miquelsthesis} M Nofrarias 2007, {\it Thermal diagnostic in the LISA technology package experiment}, Chapter 2.
PhD thesis, Barcelona.
\bibitem{nodes} A Lobo 2010, {\it LTP thermal response to periodic signals injected by heaters}, Report S2-IEC-TN-3059, v1.
\bibitem{pepsthesis} J Sanju\'an 2009, {\it Development and validation of the
thermal diagnostics instrumentation in LISA Pathfinder}, Appendix B. PhD thesis, Barcelona.
\end{thebibliography}
\end{document}